\documentclass[12pt,preprint]{aastex}
\usepackage{url}

\shorttitle{Ubiquitous Solar Eruptions Driven by Magnetized Vortex Tubes}

\title{Ubiquitous Solar Eruptions Driven by Magnetized Vortex Tubes}

\author{I.~N. Kitiashvili$^{1,2,3}$, A.~G. Kosovichev$^1$, S.~K. Lele$^{2,4}$, N.~N. Mansour$^{2,5}$, A.~A. Wray$^{2,5}$}
\affil{$^1$Hansen Experimental Physics Laboratory, Stanford University, Stanford, CA 94305, USA}
\altaffiltext{1}{e-mail: irinasun@stanford.edu}
\affil{$^2$Center for Turbulence Research, Stanford University, Stanford, CA 94305, USA}
\affil{$^3$Kazan Federal University, Kazan, 420008, Russia}
\affil{$^4$Aeronautics and Astronautics Dept., Stanford University, Stanford, CA 94305, USA}
\affil{$^5$NASA Ames Research Center, Moffett Field, Mountain View, CA 94040, USA}

\begin{document}
\begin{abstract}
The solar surface is covered by high-speed jets transporting mass and energy into the solar corona and feeding the solar wind. The most prominent of these jets have been known as spicules. However, the mechanism initiating these eruptions events is still unknown. Using realistic numerical simulations we find that small-scale eruptions are produced by ubiquitous magnetized vortex tubes generated by the Sun's turbulent convection in subsurface layers. The swirling vortex tubes (resembling tornadoes) penetrate into the solar atmosphere, capture and stretch background magnetic field, and push surrounding material up, generating quasiperiodic shocks. Our simulations reveal a complicated high-speed flow patterns, and thermodynamic and magnetic structure in the erupting vortex tubes. We found that the eruptions are initiated in the subsurface layers and are driven by the high-pressure gradients in the subphotosphere and photosphere, and by the Lorentz force in the higher atmosphere layers.
\end{abstract}

\section{Introduction}
One of the most frequently observed phenomena of dynamical coupling between the solar convection and atmospheric layers are plasma eruptions on different scales, such as CME, flares and spicules. Rapid increase of the observational power with new instruments, e.g. HMI/SDO \citep{Scherrer2012}, SOT/Hinode \citep{Tsuneta2008}, NST/BBSO \citep{Goode2010}, IMAX/SUNRISE \citep{Solanki2010} allows us to detect various features on smaller and smaller scales, ans measure their properties. Realistic numerical simulations, based on '{\it in intio}' physical principles \cite[e.g][]{stein1998,jacoutot08a,kiti2012a}, have been able to reproduce and understand physics of the observed phenomena, and predict new effects, which are hard to detect in observations. In this paper, we present new results of radiative MHD simulations, which shed light on the mechanism of small-scale eruptions in the solar atmosphere, and link it to the dynamics of turbulent magnetized vortex tubes.

Previously it was shown that an important role in dynamics the turbulent convective layers and low atmosphere, and in the chromosphere heating is played by small-scale vortex tubes, by simulations \citep{kiti2010b,kiti2011,kiti2012a,steiner2010,Moll2011} and high-resolution observations \citep[e.g.][]{Wedemeyer-Bohm2009,Balmaceda2010,Cao2010,Yurchyshyn2011,Fedun2011,Ji2012,Wedemeyer-Bohm2012}. In particular, the numerical simulations revealed the dynamics of the vortex tubes; and it was found that despite preferentially strong downflows in the vortex core, strong upflows may occur quasiperiodically, pushing matherial into the atmospheric layers (see in Fig.~5 in \cite{kiti2012a}). In this paper, we demonstrate that these quasi-periodic upflows associated with the vortex tube dynamics, in the presence of magnetic field produce small-scale jet-like ejections of plasma generating shocks in the atmosphere, and discuss the origin and properties of these eruptions.

\section{Computational setup}
To simulate the coupled dynamics of top layers of the turbulent convective zone, photosphere and low atmosphere of the Sun we use the 3D radiative MHD `SolarBox' code developed at the Stanford Center for Turbulence Research and NASA Ames Research Center \citep{jacoutot08a}. The code implements LES turbulence models, the real-gas equation of state, astrophysical opacity tables \citep{Rogers1996}, and the standard model of the solar interior for initial conditions \citep{chris1996}. Radiative transfer between fluid elements is calculated using a 3D multi-spectral-bin method, and a long-characteristic approach, assuming the local thermodynamic equilibrium.

The physical description of dynamical properties of the solar convection is improved through the implementation of subgrid-scale LES turbulence models, which can effectively increase the Reynolds number and provide representation of small-scale motions closer to the reality  \citep{jacoutot08a}. Here we used a Smagorinsky eddy-viscosity model \citep{Smagorinsky1963}, in which the compressible Reynolds stresses were calculated in the form \citep{Moin1991}: $\tau_{ij}=-2C_S\triangle^2|S|(S_{i,j}-u_{k,k}\delta_{ij}/3)+2C_C\triangle^2|S|^2\delta_{ij}/3$, where the Smagorinsky coefficients are $C_S=C_C=0.001$, $S_{ij}$ is the large-scale stress tensor, and $\triangle\equiv(dx\times dy\times dz)^{1/3}$ with $dx$, $dy$ and $dz$ being the grid-cell dimensions.

The presented simulation results are obtained for a computational domain: $6.4\times6.4\times5.5$~Mm$^3$ (with a 1-Mm layer above the photosphere, and a 5-Mm layer below) with $12.5^2\times10$~km grid-cells. We modeling the conditions of a quiet-Sun region: with an initially introduced vertical magnetic field, $B_{z0}=10$~G, and, for the comparison for pure hydrodynamic case. The lateral boundary conditions are periodic. The top boundary is open to mass, momentum and energy transfers, and also to the radiation flux. The bottom boundary is open only for radiation, and simulates the energy input from the interior of the Sun. The simulation results previously were compared with a similar type code by \cite{Nordlund2001} for some test cases, and with photospheric observations \citep{jacoutot08a,jacoutot08b,kiti2013}. Extending the computational box domain into the atmosphere allows us to model effects of the intensive energy exchange between the photosphere and the chromosphere, and investigate some of most the energetic phenomena in the quiet Sun: spontaneous high-speed flow eruptions along the magnetic flux tubes.

\section{Flow ejection due to vortex tube dynamics}

The highly-turbulent solar subsurface layers are a place of the origin of numerous vortex tubes, formed due to the granular overturning and Kelvin-Helmholtz instabilities (see Kitiashvili et al. 2012 and references therein). The turbulent vortex tubes are usually located in the intergranular lanes; they often become vertically oriented, penetrate above the solar surface, and can be stable for longer than a typical granulation life-time. High-speed, near sonic, flows of the vortex tubes are accompanying by sharp variations of temperature, density and gas pressure, and can strongly affect the dynamics of their environment. According to the previous numerical simulations, the vortex tubes penetrate into the chromospheric layers in both, magnetic and nonmagnetic cases. In the quiet-Sun region, the magnetic field effects reveal themselves mostly in the higher atmosphere as magnification of the hydrodynamic effects playing the dominant role in the turbulent surface and subsurface layers. Thus, the contribution of the magnetized vortex tubes to the chromospheric dynamics and energetics is more significant than in the hydrodynamic case \citep{kiti2012a}.

Our simulations show that the upward vortex tube penetration into the solar atmosphere is often quasi-periodic  and accompanied by spontaneous flow ejections. A time-sequence of various vortex tube properties with a cadence of 15~sec for a strong event  is illustrated in Figure~\ref{fig:time-sec}, where panels $a-~c$) show horizontal snapshots of temperature, vertical velocity and density at height $h=625$~km; and panel $d$) shows vertical cuts of log$(\rho)$ taken along the $x$-axis through the region marked by short white lines in panels $a-c$). Figure~\ref{fig:time-sec} shows a complicated structure and dynamics of the eruptions with downflows in the vortex core, and the upward eruption flows in the surrounding region. As seen in the temperature distribution plots in panel $a$) these eruptions are hotter than the surrounding plasma, and can provide an extra chromospheric heating in addition to the heating through the vortex core. The shape of the flow ejections in the vortex tubes is not completely circular due to the non-vertical vortex tube orientation, and interaction with surrounding flows (Fig.~\ref{fig:time-sec}~{\it c},~{\it d}).

The vortex tube causes strong swirling motions of the subsurface and atmospheric layers around the vortex core. Figure~\ref{fig:gradP} illustrates various stages of the flow ejection, where the streamlines show a general behavior of the velocity field; yellow-blue isosurfaces represent the pressure gradient, normalized by density, of $5\times 10^4$~cm/s$^2$ (yellow) and $-5\times 10^4$~cm/s$^2$ (blue); and the grey isosurface shows the temperature distribution of T $=6400$~K. In the atmospheric layers, the flow ejection starts with the process of formation of a vertically oriented vortex tube, as described by \cite{kiti2012b}. This creates strong vertical pressure gradients: negative in the vortex core and positive at the vortex periphery (Fig.~\ref{fig:gradP}{\it a}). The swirling motions get concentrated at a height of about 500~km (which corresponds to a temperature minimum region, panels $b$ and $c$), and then erupt (panel $d$).

Because of the strong concentrations of magnetic field in the vortex core  ($\sim 1.2$~kG in the photosphere layer) it is interesting to consider the dynamics in terms of the evolution of electric currents. Figure~\ref{fig:Electric} shows that the vortex tube penetration for the same event is accompanied by formation of an electric current sheet in the surrounding area of the intergranular lane (blue isosurface), which expands together with the vortex tube (yellow isosurface represents enstrophy) into the atmosphere and relaxes during the ejection. The current sheet can be a source of the Lorentz force (Sec.~5) and also, perhaps,  additional heating.

\section{Dynamics of flow ejections}

Generally, the flow structure during the eruption phase remains twisted: the material around the vortex core moves up from the subsurface layers, and also  towards the vortex from the surrounding region, and collects near the vortex edge. The plasma is moved up by the twisting flows into the higher atmospheric layers, and at the same time in the lower layers the plasma flows down though the vortex core (Fig.~\ref{fig:sch}{\it a}). The magnetic field lines are weakly twisted opposite to the flow direction (Fig.~\ref{fig:sch}{\it b}). Because the dynamics of eruptions is associated with the flows surrounding a relatively narrow vortex core, we track the vortex core, and analyze the data in the cylindrical coordinates centered in the vortex core. For this analysis, we choose a typical vortex tube, and divide the vortex region in 125-km 'rings' (or 'zones'), where 'zone 0' includes the vortex core and its edge  (Fig.~\ref{fig:sch1}).

Figure~\ref{fig:Vz} shows the mean vertical velocity variations with time at different heights, from $-300$~km to 780~km, for two zones: ($a$) vortex core ('zone 0'), and ($b$) surrounding region ('zone 1'). These diagrams show that the velocity perturbations are initiated at the depths of $\sim 60 - 120$~km below the surface, and then propagate in both directions, upward and downward. Such perturbations are quasiperiodic with a period of $3 - 5$~min, that can be explained by a similar characteristic turnover time of the turbulent behavior of convection, where vortex tube rooted. The period is also similar to the oscillations of large-scale acoustic ($p$) modes excited in the box domain. However, these oscillations do not show correlation in phase with the vortex-tube oscillations, and their amplitude is essentially smaller. The mean vertical velocity of the perturbations increases up to 5 -- 8~km/s at $\sim 800$~km  above the solar surface. These upflows can be identified as small-scale flow ejections, driven by vortex tubes. The amplitude  and quasi-period of the eruptions vary due to the vortex evolution (e.g., changes of their size, shape, height penetration) or/and interactions with other vortices, as a result of which the  ejections can be magnified or suppressed.

The comparison of the vertical velocity variations (Fig.~\ref{fig:Vz}) in the vortex core (panel $a$) and the surrounding region (panel $b$, as indicated in Figure~\ref{fig:sch1} zones '0' and '1'), shows that in the vortex core the velocity perturbations have stronger amplitude. In the core, the upward speed of the velocity perturbations increases with height from 6~km/s in the near-surface layers to more than 12~km/s above 700~km. The downward perturbations propagate much slower, with a speed of $3-3.5$~km/s, and their amplitude increases.
In the vortex surrounding region (Fig.~\ref{fig:Vz}{\it b}), the velocity shows a similar behavior. The time shift of the vertical velocity variations between different zones allows us to estimate the speed of the flow expansion during the eruption, which is about 20 -- 25~km/s in the vortex core area and decelerates to $\sim 15$~km/s at a distance of about 500~km from the vortex. For each individual ejection event, the estimate can vary due to the interaction of the perturbations with a shocks from other eruptions.

The transformation of the velocity perturbations into shock waves additional interesting feature, associated with the flow ejection is  (Fig.~\ref{fig:time-sec}{\it c, d}). This effect is identified in both the hydrodynamic and magnetic simulations, and in both cases associated with the vortex tube dynamics. In the next section we consider in more detail the process of the flow ejection and magnetic field effects.

\section{Source and drivers of spontaneous flow eruptions}

The complicated dynamics of the strong swirling flows in the presence of magnetic field across many pressure scale-heights is an interesting interplay of hydro- and magnetic effects. In general, there are two type of forces that are responsible for driving the flow  eruptions: hydrodynamic, due to pressure excess, and magnetic, caused by the Lorentz force. A comparison of the contributions from the hydrodynamic and magnetic effects can be done by estimating the RHS terms of a modified momentum equation:
\begin{equation}
    \frac{\partial{\rm\bf v}}{\partial t}+({\rm\bf v}\cdot\nabla){\rm\bf v}=\frac{{\rm\bf J}\times{\rm\bf B}}{c\rho}-\frac{\nabla p}{\rho}-g, \nonumber
\end{equation}
where {\bf v} is the velocity vector, {\bf J} is the electric current density, {\bf B} is the magnetic field vector, $p$ is the gas pressure, $\rho$ is the density, $g$ and $c$ are the gravity acceleration and the speed of light. In this form, Eq. (1) describes the flow acceleration on the left hand side, and the contributions of the Lorentz force (first term) and the non-magnetic forces: the pressure gradient and gravity. In the initial equilibrium state the pressure gradient and gravity are balanced.

Figures~\ref{fig:gradP-prof} and~\ref{fig:FL-prof} show a comparison of the pressure gradient excess and the Lorentz force in the vortex core (zone '0', panels $a$) and the surrounding region (zone '1', panels $b$). The profiles of both, non-magnetic and magnetic forces, reveal a clear connection with the upward and downward velocity perturbations, associated with the flow ejection, showing a similar decrease of the perturbation amplitude in the vortex surrounding regions, and also the time lag with height. Propagation of the perturbations is better visible in the Lorentz force (Fig.~\ref{fig:FL-prof}) with a clear indication of acceleration in higher layers of the atmosphere, indicating a strong increase of magnetic field effects at the height $h\geq 700$~km. The propagation of the Lorentz force perturbations upward and downward along the vortex tube (Fig.~\ref{fig:FL-prof}{\it a}) gives us an additional estimation for the depth of the initialization source, which varies among different events from the photosphere to $\sim 120$~km below surface. This corresponds to the non-magnetic case, where the primary source of the vortex eruptions is hydrodynamic. Magnetic field effects become important above the temperature minimum region, where upflow are accelerated by the Lorentz force.

Figure~\ref{fig:accel-prof} shows the contributions of the magnetic (blue curves) and nonmagnetic acceleration (red curves) in four layers: 780~km and 300~km above the solar surface (panels {\it a} and {\it b}), the photosphere layer (panel {\it c}), and 240~km below the surface (panel {\it d}). Dashed curves represent the vertical velocity for the same layers, given for comparison. The results show that the Lorentz force is most important for the flow acceleration in the higher layers, where strong Lorentz-force fluctuations are correlated with the strong flow acceleration (e.g. at $t=9$~min). The flow eruption is much weaker for the events when most contribution in these layers comes from the hydrodynamic force (e.g. $t=3$~min). Nevertheless, close to the surface and in the subsurface layers the hydrodynamic effects are significantly more important (Fig.~\ref{fig:accel-prof}~{\it b-d}) than the Lorentz force.

As we have discussed above, the initial perturbations of the flow velocity and the hydro- and magnetic forces associated with the vortex eruption are generated just below the surface, where the effect of the pressure gradient force is dominant. The subsurface layers in the vicinity of the vortex core are characterized by strong converging flows, which compress the vortex tube, magnify the upward pressure gradient and accelerate the vortex-tube swirling motions. Figure~\ref{fig:time-sec-300} illustrates the compression process in a layer located 300~km below the surface (panel {\it a}), and the evolution of the vortex at the surface (panel {\it b}) with a 30-sec cadence. The middle snapshots in both panels correspond to the first snapshot in the atmosphere layer at the height of 625~km, presented in Figure~\ref{fig:time-sec}, and illustrate the beginning of the eruption. Such compression in the subsurface layers is continuous and nonstationary process, which is affected by convective oscillations and local turbulent magnetohydrodynamics.

\section{Conclusion}

\begin{figure}
\begin{center}
\includegraphics[scale=1.5]{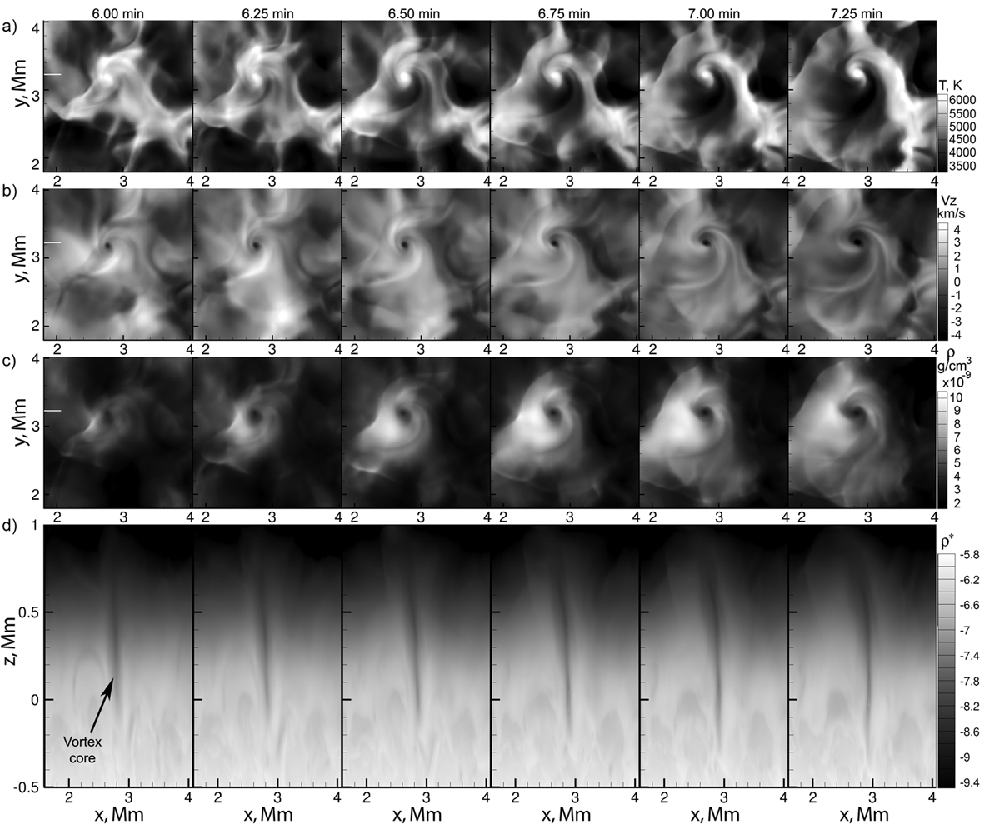}
\end{center}
\caption{Time-series with a cadence of 15~sec of temperature (panel {\it a}), vertical velocity ({\it b}) and density ({\it c}) at the height of 625~km above the solar surface; and ($d$) a vertical slice for the density, $\rho^*=$log$(\rho)$) for a flow ejection initiated by a vortex tube. The location of the vertical slice is indicated by a short white line on the horizontal images. Time starts from the moment of the beginning of the vortex tube tracking for the analysis in Figures~\ref{fig:Vz} -~\ref{fig:accel-prof}\label{fig:time-sec} }
\end{figure}

\begin{figure}
\begin{center}
\includegraphics[scale=0.8]{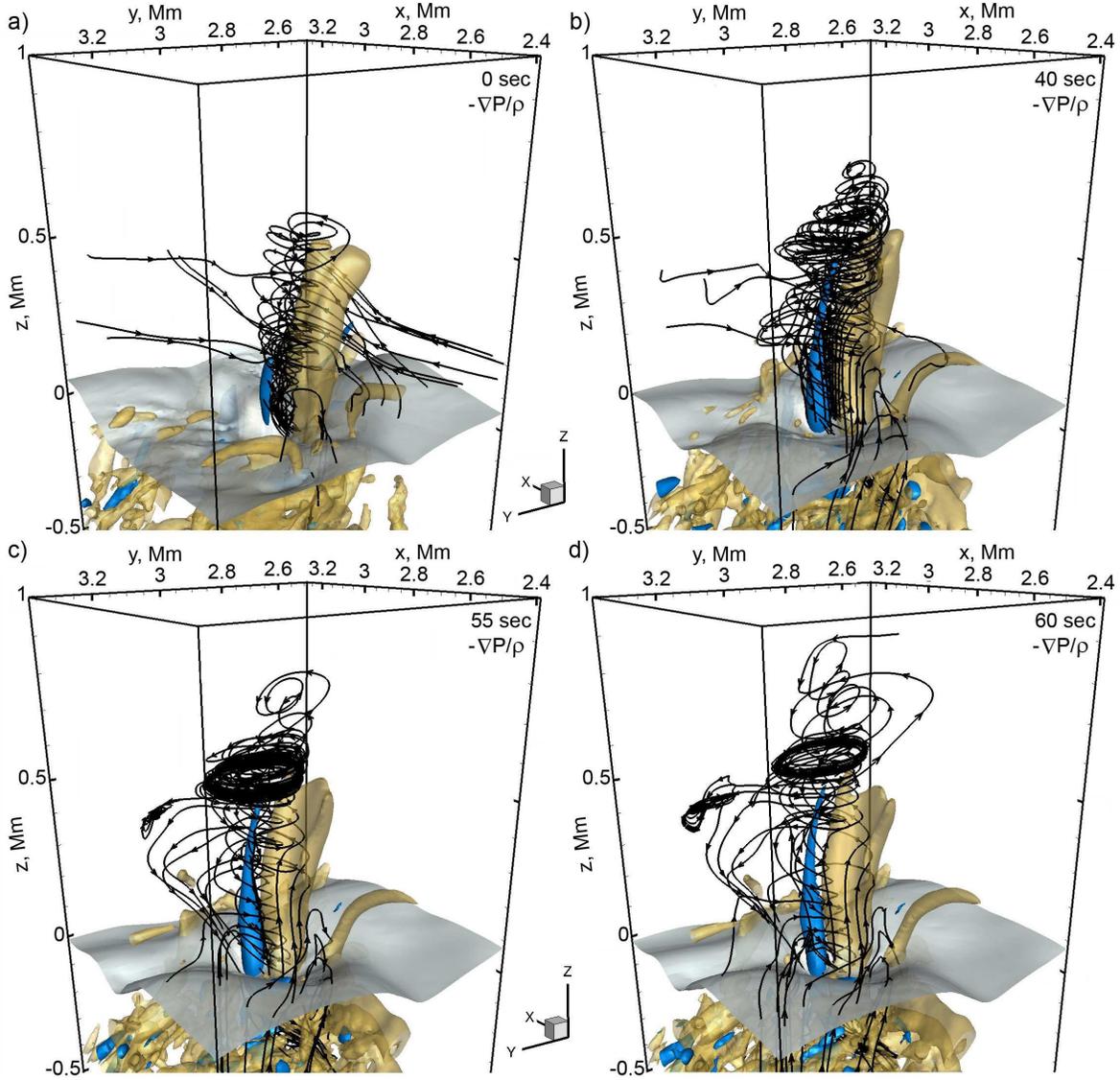}
\end{center}
\caption{Different stages of the flow ejection: {\it a}) vortex tube penetration into the atmosphere layers, {\it b}) intensification of the swirling motions, {\it c}) concentration of the swirling motion in a ring-like structure, and {\it d}) flow ejection along the vortex tube. Black streamlines illustrate the velocity field in the vicinity of the vortex tube. Semitransparent light grey surface corresponds to a constant temperature of 6400~K. Yellow and blue isosurfaces correspond to the normalized-by-density vertical pressure gradient ($-\nabla p/\rho$) for the values of $5\times10^4$~cm/s$^2$ (yellow color) and $-5\times10^4$~cm/s$^2$ (blue color). \label{fig:gradP}}
\end{figure}

High-resolution ground-based and space observations have revealed an intense and very dynamic interaction between the surface layers and the low atmosphere in quiet-Sun regions with relatively weak mean magnetic field. The radiative MHD simulations can reproduce some basic features of observed phenomena and provide an important complimentary tool for investigation of the underlying physical processes.
Our numerical simulations with an initial mean 10~G vertical magnetic field show a complicated mixture of hydrodynamic and magnetic effects associated with spontaneous quasi-periodic (with period 3 -- 5~min) flow ejections from the subsurface layers into the higher atmosphere along the magnetized vortex tubes, in which the magnetic field strength on the surface is typically $\sim 1$~kG (Fig.~\ref{fig:Vz}). The eruptions have a complicated dynamical structure with mostly continuous swirling downflows, decreasing density and heating in the vortex core, and spontaneous upflows mostly propagating along the vortex core periphery (Figs.~\ref{fig:time-sec} and~\ref{fig:gradP}) and forming shock waves in the higher atmosphere (Figs.~\ref{fig:time-sec}{\it c} and {\it d}). The plasma flow in the eruptions accelerate in the higher (mid-chromospheric) layers from 6 to $12 - 15$~km/s. Also, the perturbations, associated with the flow ejection, propagate into the solar interior along the vortex tube core with a speed about $3 - 3.5$~km/s and increasing amplitude.

\begin{figure}
\begin{center}
\includegraphics[scale=0.65]{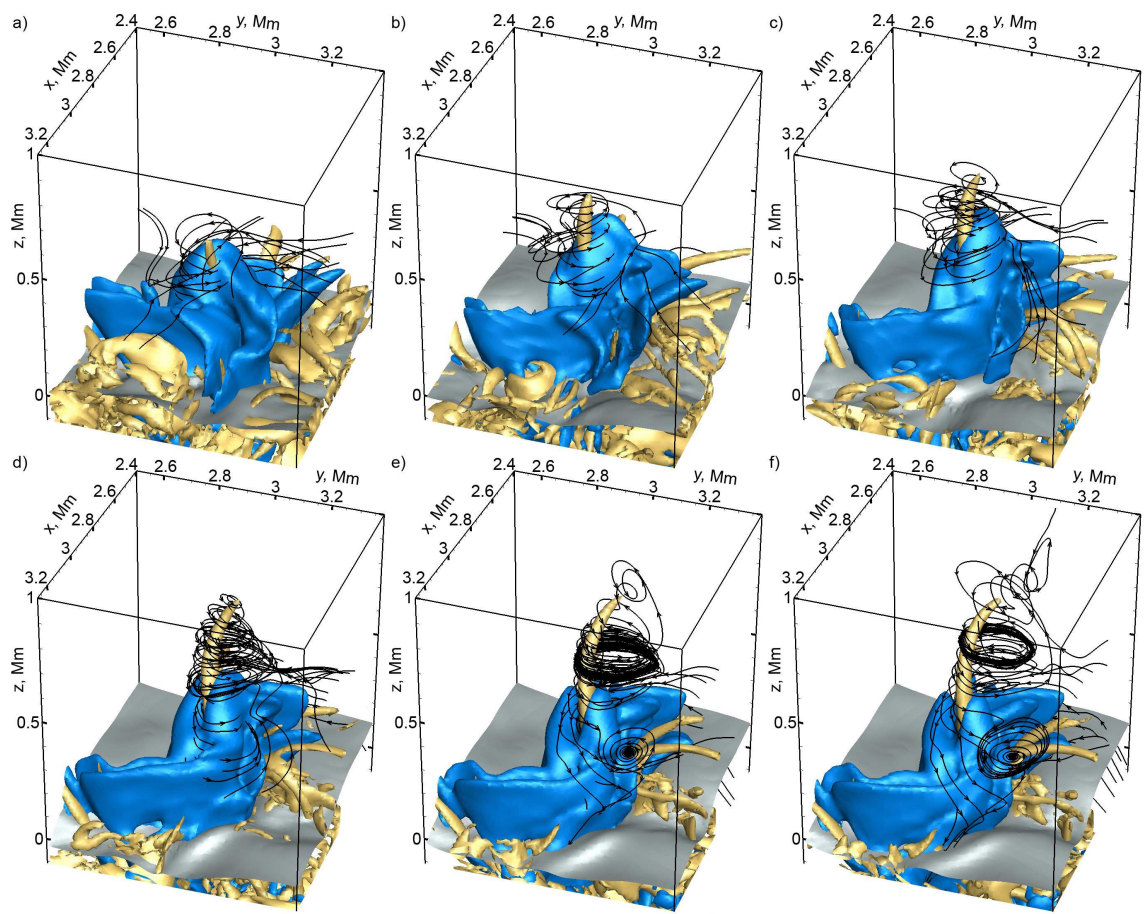}
\end{center}
\caption{Evolution of the electric current sheet during the vortex tube eruption, which drags it into the higher atmospheric layers (panels {\it a-d}) followed by the current relaxation during the flow ejection (panels {\it e-f}). Blue isosurfaces correspond to the value of electric current $|J|=8\times 10^4$ and yellow isosurfaces show the enstrophy distribution for $\Omega =|\textrm{curl}${\bf v}$|^2= 0.35$~cm$^{-2}$. Black streamlines show selected flow trajectories. \label{fig:Electric} }
\end{figure}

\begin{figure}
\begin{center}
\includegraphics[scale=1.4]{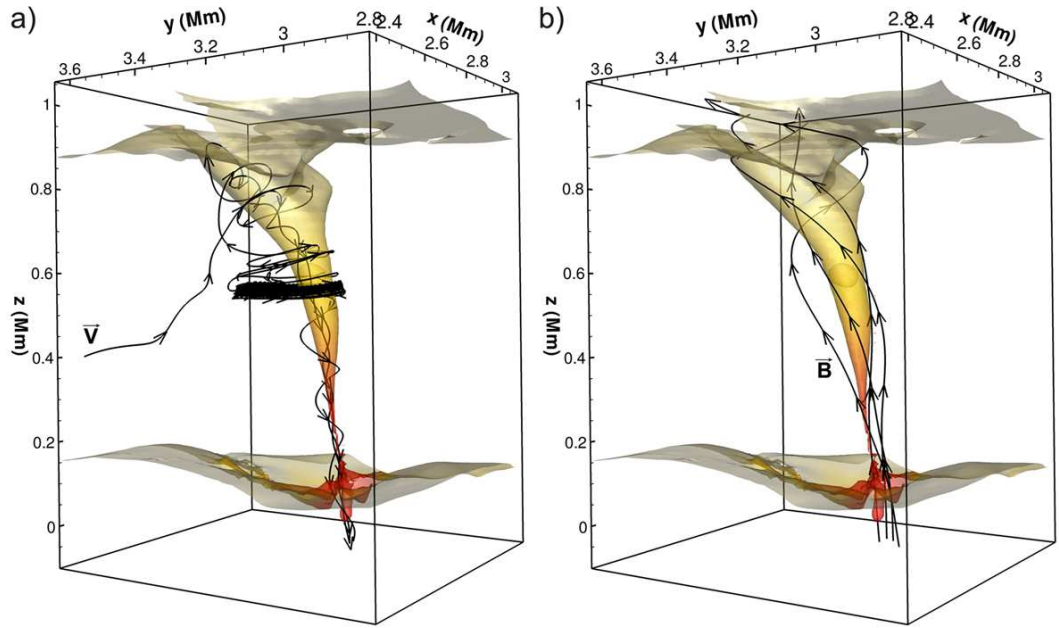}
\end{center}
\caption{Topology of the flow streamlines and magnetic field lines during the flow ejection. The vortex is visualized by a constant temperature isosurface, $T=6000$~K. The color-scale on this surface shows the distribution of the magnetic field strength from 50~G or less (grey) to 1.2~kG (red). Streamlines in panel $a$) illustrate the topology of flows, with helical upward ejection flows and downflows in the narrow vortex core. In panel $b$) black lines show the topology of the magnetic field lines. \label{fig:sch} }
\end{figure}

\begin{figure}
\begin{center}
\includegraphics[scale=0.8]{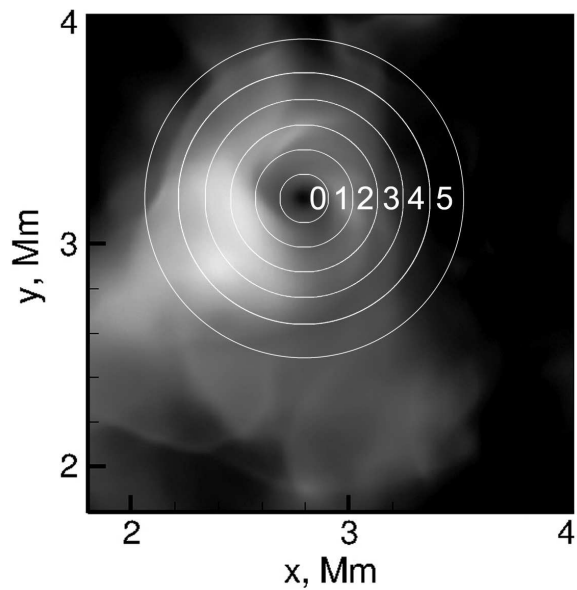}
\end{center}
\caption{Schematic representation of the analysis on zones with 125~km width around the vortex core. Background image illustrates the density distribution at height of 625~km above the solar surface for $t=7$~min. For the analysis, presented in Figures~\ref{fig:Vz} -~\ref{fig:accel-prof}, the vortex core tracked in time for the subsurface and atmosphere layers. \label{fig:sch1} }
\end{figure}

\begin{figure}
\begin{center}
\includegraphics[scale=0.76]{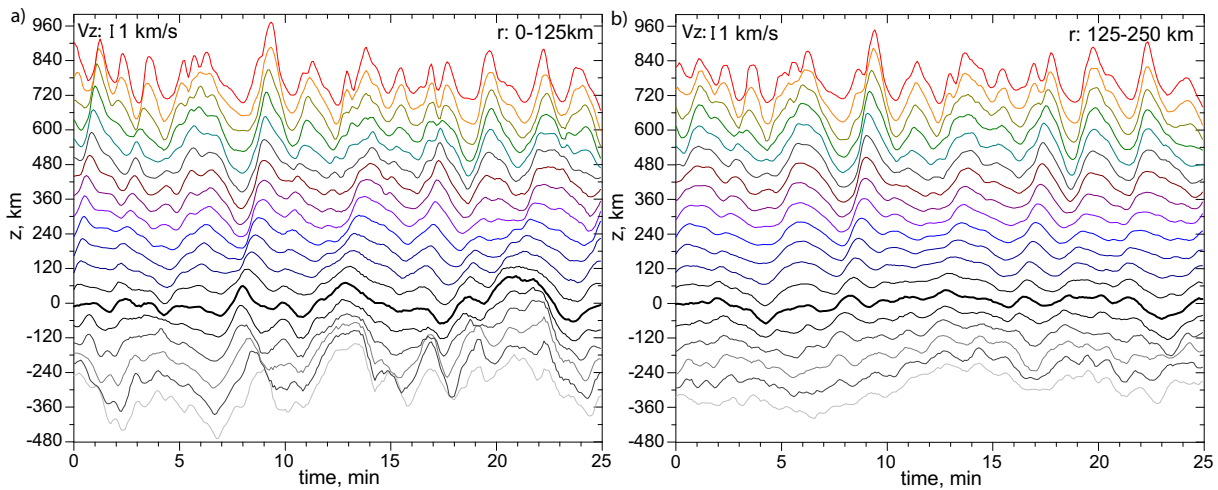}
\end{center}
\caption{Temporal profiles of variations of the mean vertical velocity, $V_z$ in: $a$) vortex core, region '0' (range 0 -- 125~km) and $b$) surrounding region '1' (range 125 -- 250~km) at different levels below the surface and in the atmosphere. Thick black curve shows the variations in the photosphere layer. The height difference between the curves is 60~km.  \label{fig:Vz} }
\end{figure}

\begin{figure}
\begin{center}
\includegraphics[scale=0.76]{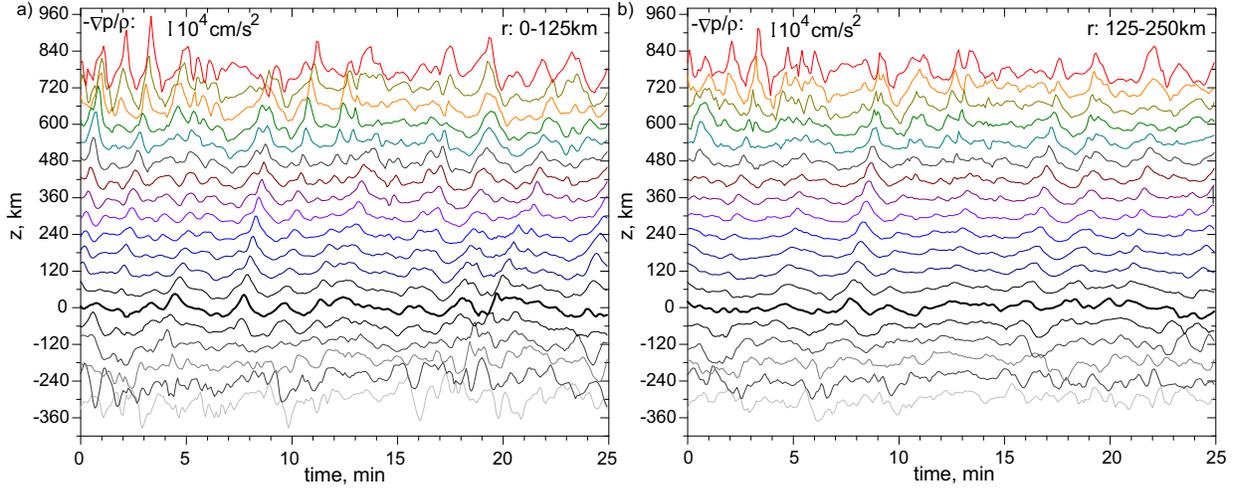}
\end{center}
\caption{Temporal profiles of mean of the pressure gradient variations for: $a$) vortex core region '0' (range 0 -- 125~km) and $b$) surrounding region '1' (range 125 -- 250~km) at different levels below the surface and in the atmosphere. Thick black curve shows the variations in the photosphere layer. The height difference between the curves is 60~km.  \label{fig:gradP-prof} }
\end{figure}

\begin{figure}
\begin{center}
\includegraphics[scale=0.76]{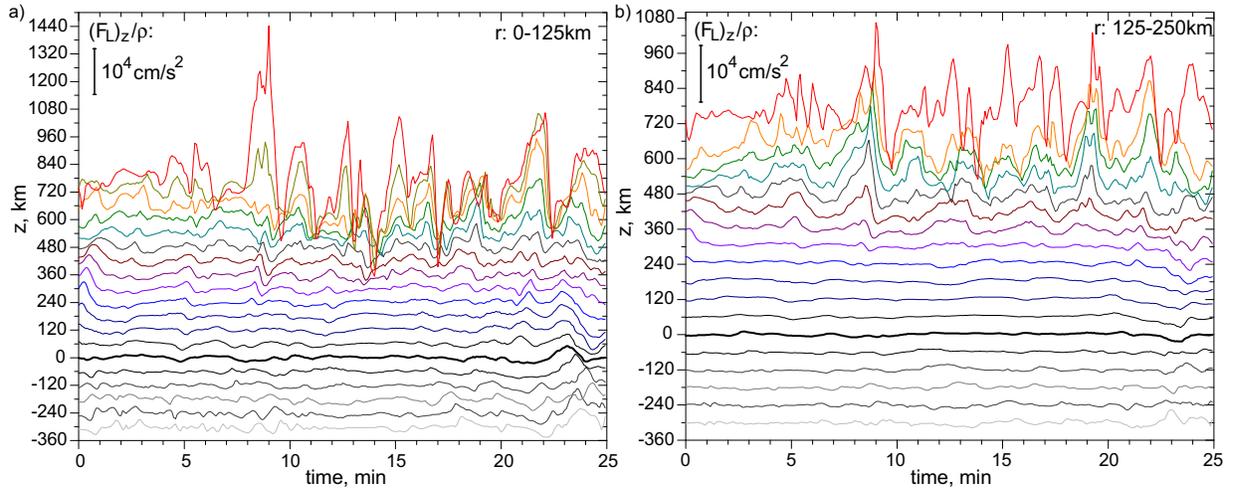}
\end{center}
\caption{Temporal profiles of the mean variations of the Lorentz force normalized by density $(F_{L})_z/\rho$, for: $a$) vortex core region '0' (range 0 -- 125~km) and $b$) surrounding region '1' (range 125 -- 250~km) at different levels below the surface and in the atmosphere. Thick black curve shows the variations in the photosphere layer. The height difference between the curves is 60~km.   \label{fig:FL-prof} }
\end{figure}

\begin{figure}
\begin{center}
\includegraphics[scale=0.95]{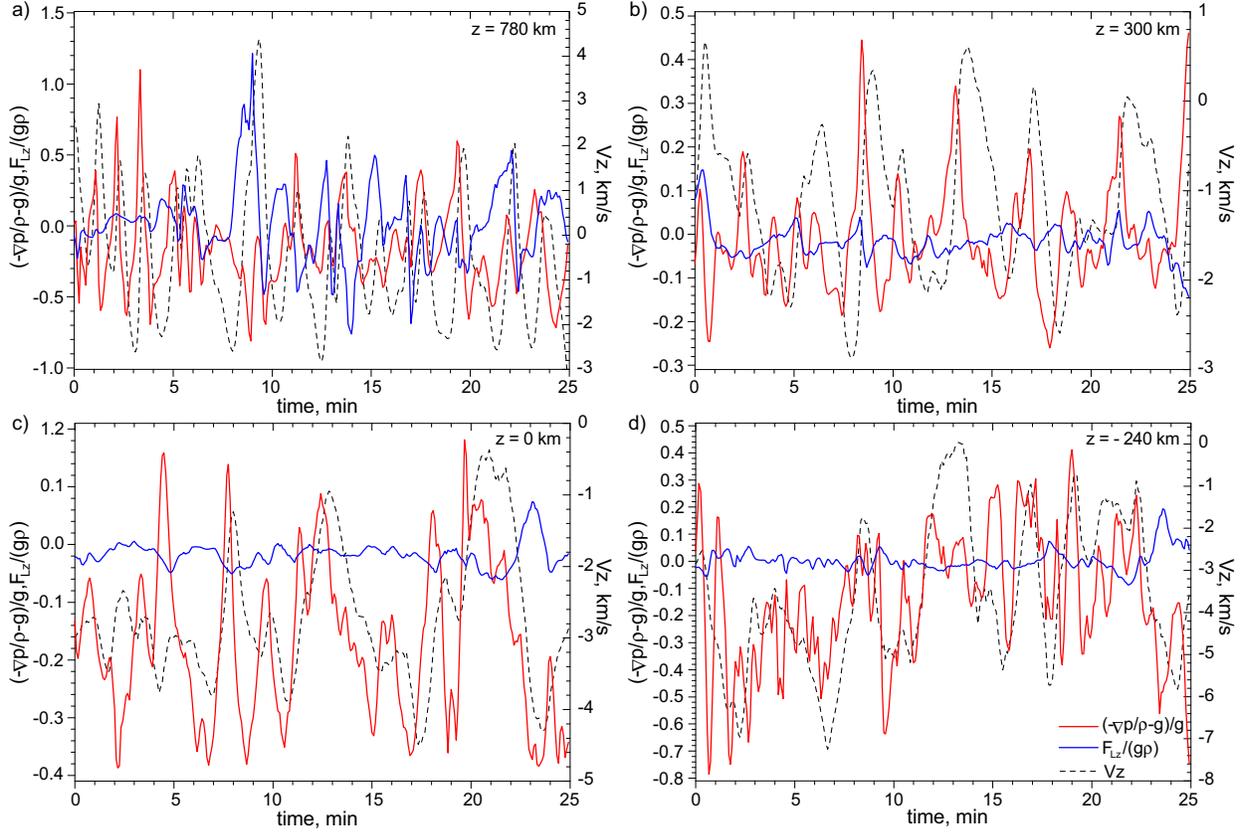}
\end{center}
\caption{Comparison of the contributions to the flow ejection of the hydrodynamic ($(-\nabla p/\rho-g)/g$, red curves) and magnetic (Lorentz force, $F_L/(\rho g)$, blue curves) vertical accelerations (normalized by the gravity acceleration) at different altitudes: {\it a}) $h=780$~km, {\it b}) $h=300$~km, {\it c}) $h=0$~km (photosphere) and {\it d}) $h=-240$~km. Dashed curves correspond to the vertical velocity, $V_z$, given for reference at the same layers. All curves correspond to the vortex core region '0'. \label{fig:accel-prof} }
\end{figure}

\begin{figure}
\begin{center}
\includegraphics[scale=1.5]{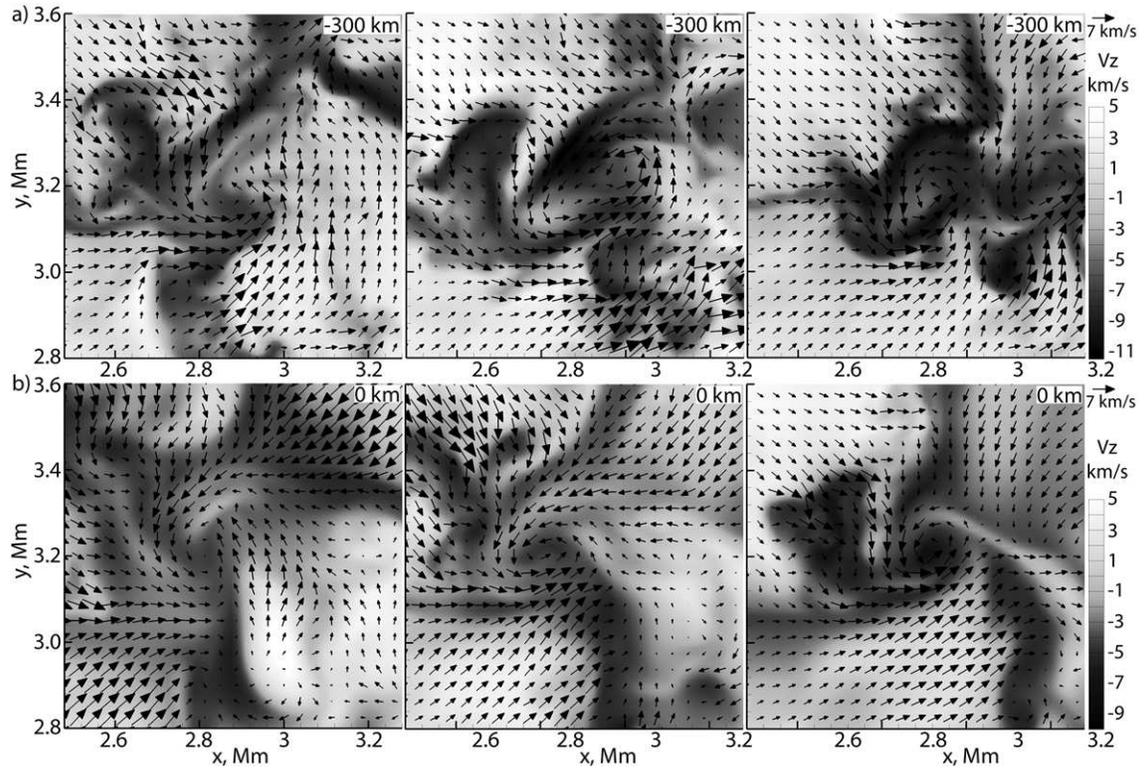}
\end{center}
\caption{Compression of the vortex region by surrounding converging flows. Background images show the vertical velocity distribution at the depth of 300~km (panel{\it a}) and in the photosphere layer (panel {\it b}) with cadence 30sec. Arrow show the horizontal velocity field. Middle snapshots in each panel correspond to the first snapshot in Figure~\ref{fig:time-sec}.\label{fig:time-sec-300} }
\end{figure}

The process of the flow ejection originates in a subsurface 100~km  deep layer, where the vortex compression by converging flows increases the pressure gradient and accelerates the swirling flows (the evolution of a strong event is shown in Figs.~\ref{fig:time-sec} -- \ref{fig:Electric}). This compressed vortex tube starts penetrating into the low-atmosphere layers, and involves the surrounding atmospheric plasma into swirling motion. Accumulation of the swirling flows at a height of $\sim 500$~km (a temperature minimum region) forms an ring-like structure, which propels the swirling flows into the higher layers along the vortex tube. The flows are mostly accelerated by the pressure gradient in the subsurface and near-surface layers, and by the Lorentz force in the higher layers ($\geq$ 700~km).
The described mechanism of the flow ejections in the vortex tubes works also in the pure hydrodynamic case, but in this case the velocity perturbations are much smaller and, the eruptions almost immediately fall back to the photosphere. Nevertheless, in this case the vortex tube eruptions also generate shocks in the low-density atmospheric layers.

Thus, we can conclude that the magnetic field captured in the vortex tubes plays an important role, magnifying the initial hydrodynamic perturbations, and accelerating them along the vortex tubes by the Lorentz force  in the higher atmosphere, producing ubiquitous spontaneous flow eruptions. Our next step is to investigate the propagation of these eruptions into the solar corona.

\end{document}